# *In situ* time-resolved X-ray absorption spectroscopy of shock-loaded magnesiosiderite

Anand Prashant Dwivedi[1,2*], Jean-Alexis Hernandez[3], Sofia Balugani[3], Delphine Cabaret[4], Valerio Cerantola[5], Davide Comboni[6], Damien Deldicque[7], François Guyot[4], Marion Harmand[4], Harald Müller[3], Nicolas Sévelin-Radiguet[3], Irina Snigireva[3], Raffaella Torchio[3], Tommaso Vinci[8], Thibaut de Rességuier[9]

[1*]European X-ray Free-Electron Laser Facility, Schenefeld, Germany.

[2*]Department of Physics, Universität Hamburg, Hamburg, Germany.

[3]European Synchrotron Radiation Facility, Grenoble, France.

[4]Institut de Minéralogie, de Physique des Matériaux et de Cosmochimie (IMPMC), UMR7590, CNRS, Sorbonne Université, Muséum National d'Histoire Naturelle, Paris, France.

[5]Department of Earth and Environmental Sciences, University of Milano-Bicocca, Milano, Italy.

[6]Department of Earth Sciences, University of Milano, Milano, Italy.

[7]Laboratoire de Géologie, Département de Géosciences, École Normale Supérieure, CNRS, UMR 8538, PSL University, Paris, France , Paris, France.

[8]Laboratoire pour l'Utilisation des Lasers Intenses, École Polytechnique, Palaiseau, France.

[9]Institut PPRIME, CNRS-ENSMA-Université de Poitiers, Poitiers, France.

*Corresponding author (anand.dwivedi@xfel.eu)

# Abstract

Carbonate minerals are important in Earth's system sciences and have been found on Mars and in meteorites and asteroids, highlighting the importance of impacts in planetary processes. While extensively studied under static compression, the behavior of carbonates under shock compression remains underexplored, with no *in situ* X-ray investigations reported so far. Here we investigate natural magnesiosiderite ($Fe_{0.6}Mg_{0.4}CO_3$) under nanosecond laser-driven shock compression at pressures up to 150 GPa, coupled with *in situ* ultrafast synchrotron X-ray absorption spectroscopy (XAS). The interpretation of the experimental spectra is complemented using first-principles absorption cross-section calculations performed on crystalline phases at different pressures and on a dense liquid phase obtained using density functional theory-based molecular dynamics (DFT-MD) simulations. Under laser-driven shock compression, the magnesiosiderite crystal phase remains unchanged up to the melt. Under shock reverberation, the absorption spectra show changes similar to those attributed to a high-spin to low-spin transition observed under static compression. At higher pressures, the laser shock induces the formation of $CO_4$ tetrahedral units in the melt. Upon unloading from the shocked state, only a few nanoseconds later, the original magnesiosiderite phase is recovered.

## I.   INTRODUCTION

Carbonate minerals are among the most widely distributed minerals in the Earth's crust and serve as the planet's main carbon reservoir. They play a dominant role in Earth's deep carbon cycle, including the transfer of carbon from the crust to the deep mantle through subduction processes [1–3]. In recent decades, carbonates have also been identified in various extraterrestrial environments. Outcrops rich in iron-magnesium carbonates have been identified in the Columbia Hills of Gusev crater on Mars by the Spirit rover [4]. Carbonates were also detected on asteroid Bennu via spectroscopic observations [5] and in the returned samples from asteroid Ryugu [6]. Additionally, carbonates were identified in the Martian meteorite ALH84001 [7], where they are argued to be the source of magnetite ($Fe_3O_4$), formed as a result of shock impact-driven decomposition of iron carbonates [8, 9].

Consequently, significant efforts have been devoted to characterizing the properties of carbonates, especially siderite ($FeCO_3$) and its solid solution with magnesite ($MgCO_3$), using static compression techniques [10–28].

High static pressures induce multiple changes in siderite, with the most prominent being the spin crossover from the high-spin state to the low-spin state, which occurs at ~45 GPa at room temperature [11, 15, 18], and is shifted to ~53 GPa at elevated temperatures of ~1200 K [16]. The pressure-induced spin transition in siderite significantly alters its properties, leading to, for example, a reduction in cell volume [11, 15, 29]. At pressures above 74 GPa and temperatures greater than 1750 K, corresponding to the conditions in Earth's lower mantle at depths greater than 1800 km, siderite and ferromagnesian carbonates transform to a phase with tetrahedrally coordinated carbon [17, 22, 30]. Upon reaching the melt under static pressures, siderite partially decomposes into iron oxides, with carbon and oxygen as by-products [19, 22, 27].

While iron-rich carbonates have been extensively studied under static compression, their investigation under shock compression remains limited. Recently, the first shock Hugoniot of an iron-rich carbonate, $(Fe_{0.75}Mg_{0.25})CO_3$, has been measured up to ~90 GPa and ~1700 K using a two-stage light gas gun setup [31]. Two previous shock recovery studies have examined natural siderite crystals under shock-loading conditions [32, 33], primarily focusing on the impact-driven decomposition of siderite to magnetite, relevant to the Martian meteorite ALH84001. After nanosecond laser-driven shock compression to a pressure of ~13 GPa, a magnetite-like phase was detected in the recovered sample using Raman spectroscopy and transmission electron microscopy (TEM) [32]. For shock pressures above 39 GPa generated in natural siderite by impact-driven projectiles [33], magnetic particles were detected in the recovered samples using a hand magnet. Magnetite particles, ~50-100 nm wide, were identified using TEM at shock pressures up to 49 GPa. In both experiments [32, 33], analysis was conducted on recovered samples, leaving many open questions that would benefit from *in situ* investigation. Key questions include: What crystallographic phases and electronic transitions are observed *in situ* under shock compression? Does the electronic spin transition observed under static compression also occur during shock compression? Does magnetite form upon shock-induced decomposition of iron carbonates? How does the high-pressure behavior of siderite and magnesiosiderite differ between static and dynamic compression? What is the composition of the melt induced by shock loading? To address these questions, we have performed nanosecond laser-driven shock compression on natural magnesiosiderite crystals, coupled with *in situ* ultrafast synchrotron X-ray absorption spectroscopy (XAS).

XAS is a widely used technique for determining the local electronic and structural properties of matter. It is sensitive to the oxidation and spin states, and provides information about the local atomic structure and disorder [34]. In recent years, with the advent of laser-driven shock-compression setups at large-scale X-ray facilities, XAS has been used as an *in situ* diagnostic during shock compression over extremely short time scales, with accurate pump-probe synchronization [35–39].

In this study, we report *in situ* observations of laser shock-compressed natural magnesiosiderite under shock loading and unloading, up to the melt at pressures above 100 GPa. Our experimental results are complemented using first-principles calculations and molecular dynamics simulations. During shock loading, the XAS spectra show changes similar to those attributed to the electronic spin transition under static compression [18], and suggest a kinetic- and/or thermodynamic-dependent nature of the spin transition. Contrary to the crystalline phase transition to tetracarbonates observed under static compression [22, 40], our molecular dynamics simulations indicate the formation of $CO_4$ tetrahedral units in the shock-induced melt. Upon release from the shocked state, the XAS spectra suggest recrystallization of magnesiosiderite within a few nanoseconds, independent of the peak shock pressure. Post-impact recovered samples indicate no decomposition to magnetite.

## II. METHODS

### A. Sample

The studied crystals are natural magnesiosiderites collected at the dumps of the Mésage mine, Saint Pierre de Mésage (Isère, French Alps). The crystals were selected using Raman spectroscopy based on the criterion that the obtained Raman spectra (e.g., Fig. 1(e)) are consistent with those of a FeCO$_3$ – MgCO$_3$ solid solution with varying Fe and Mg concentrations towards the iron endmember [41]. The crystals were double-polished down to a thickness of 20-21 µm to ensure optimal XAS measurements under laser shock compression. Scanning Electron Microscopy/Energy Dispersive X-ray spectrometry (SEM/EDX) showed that the samples are chemically homogeneous, containing only Fe, C, O, and Mg, with Mn and Ca contents less than 0.5 wt%. Single-crystal X-ray diffraction (XRD) measurements were performed at the ID15-B ESRF beamline [42] at ambient conditions, utilizing a convergent monochromatic beam of ~30 keV energy ($\lambda$ = 0.41 Å). The sample-to-detector distance was calibrated using a Si standard and a vanadinite (Pb$_5$(VO$_4$)$_3$Cl) single crystal. The single-crystal XRD of the magnesiosiderite crystal yields the lattice parameters a = b = 4.6698 Å, c = 15.2455 Å (space group $R\bar{3}c$), and volumic mass $\rho_0$ = 3.573 g cm$^{-3}$. Structural refinement indicates an occupancy of 0.6 for Fe and 0.4 for Mg in the (6$b$) Wyckoff position, giving the Fe$_{0.6}$Mg$_{0.4}$CO$_3$ composition for our sample (referred to as magnesiosiderite), consistent with SEM/EDX analysis.

B. Experimental setup

*In situ* XAS measurements under laser-driven shock compression were performed at the High Power Laser Facility on the ID24-ED beamline at the ESRF [43, 44]. XAS measurements were performed at the Fe *K*-edge (7.112 keV) using a 100 ps long X-ray pulse. To generate the shock, a 1053 nm, up to 40 J energy, 10 ns duration laser pulse was focused onto a spot of 250 $\mu$m or 100 $\mu$m diameter, leading to effective intensities of up to 4 × 10$^{12}$ W cm$^{-2}$. The drive laser and the X-rays were at angles of 15 degrees and -15 degrees, respectively, from the normal axis of the target plane. By varying the delay between the drive laser and the X-ray probe, the absorption spectra were collected at different times
relative to the shock breakout from the target.

The targets are multilayered (Fig. 1(a)), with aluminium-flash-coated natural crystals of Fe$_{0.6}$Mg$_{0.4}$CO$_3$ glued onto either 70 $\mu$m thick parylene or 75 $\mu$m thick black Kapton acting as ablators. Upon laser irradiation, the ablator was vaporized into plasma, and its expansion towards the laser source induced, by reaction, a compression wave in the solid sample. For a few shots, an additional rear layer of 30-35 $\mu$m thick sapphire was glued on top of the Fe$_{0.6}$Mg$_{0.4}$CO$_3$ crystals. This sapphire rear window enabled increasing both the pressure and its duration of application, thanks to a shock being reverberated at the magnesiosiderite - sapphire interface. A line-imaging Velocity Interferometer System for Any Reflector (VISAR) [45] was used as a diagnostic for time-resolved measurements of the mean shock velocity across the sample thickness, the ablator-sample interface velocity, the sample's free surface velocity, or the sample-sapphire interface velocity, depending on the target setup. Note that magnesiosiderite was found to be partially transparent to the 532 nm wavelength of the VISAR probe laser, so both the break-in time of the shock wave entering the magnesiosiderite layer and the break-out time of the wave exiting were probed simultaneously.

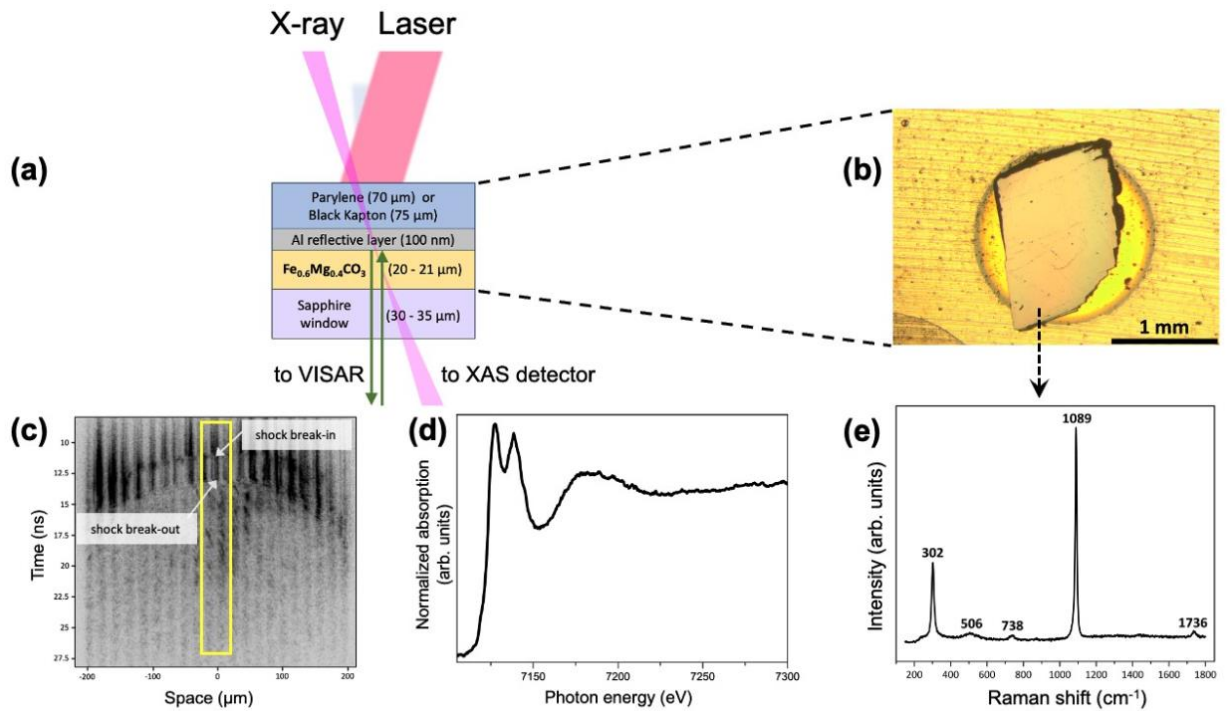

FIG. 1. Schematic of the experimental setup. **(a)** Configuration of the multi-layered sample and experimental setup, with the drive laser and X-ray probe at angles of 15 degrees and -15 degrees from the normal axis of the target plane respectively. A Velocity Interferometer System for Any Reflector (VISAR) system and a single pulse of X-rays of 100 ps duration probe the shocked state in the sample. **(b)** Top view of a polished natural magnesiosiderite crystal glued on a parylene ablator. **(c)** Example of a VISAR record showing the shock break-in and break-out times in the magnesiosiderite crystal. The region of interest of the X-ray probe area is shown in yellow. **(d)** The corresponding X-ray absorption spectrum of the shock-loaded sample probed using a single pulse of X-rays of 100 ps duration. **(e)** Corresponding Raman spectrum of natural magnesiosiderite Fe$_{0.6}$Mg$_{0.4}$CO$_3$.

## C. Determination of shock pressure

Shock pressure $P$ is obtained using the following Rankine-Hugoniot conservation relation [46]:

$$P = \rho_0 U_s u_p \qquad (1)$$

where $\rho_0$ = 3.573 g cm$^{-3}$ is the initial density of the sample, determined from single-crystal X-ray diffraction structural refinement. $u_p$ is the particle velocity, approximated by $U_{fs}$ = $2u_p$ [47, 48], where $U_{fs}$ is the free surface velocity measured from the VISAR signal at the interface between the sample and vacuum. This approximation has been shown to be valid for shocked magnesiosiderite by comparison with impedance matching measurements [31]. For higher pressure shots, and for shots where a clear free surface velocity could not be measured, a linear fit to the $u_p$ vs. effective laser intensity plot (Supplementary Fig. 7) is used to infer the particle velocity $u_p$. The shock velocity, $U_s$, is then determined using the following Hugoniot relation:

$$U_s = 5.96 + 0.71 u_p \qquad (2)$$

which is obtained using a linear fit to the $U_s$-$u_p$ relation from the Hugoniot data of (Fe$_{0.75}$Mg$_{0.25}$)CO$_3$ reported by Wang et al. (Supplementary Fig. 8) [31].

Supplementary Fig. 1 shows the impedance matching construction for pressure determination from VISAR data. The free surface velocity of magnesiosiderite is measured when no additional rear layer of sapphire is used (Supplementary Fig. 1(a,b)), and the pressure is calculated using equations (1) and (2), assuming no significant pressure decay during propagation across the ~20µm thick sample. When an additional rear layer of sapphire is used behind the magnesiosiderite crystal, the impedance mismatch between both layers is accounted for to determine the shocked state in magnesiosiderite before the shock wave breaks out in sapphire (Supplementary Fig. 1(c,d)). The single adiabat approximation is used to relate the particle velocity and pressure in magnesiosiderite ($u_{p1}$, $P_1$) with those measured in sapphire ($u_{p2}$, $P_2$). As an example, and as a result of the shock reverberation at the magnesiosiderite - sapphire interface, the pressure increases from ~56 GPa before break-out to ~68 GPa after re-shock for the data presented in Supplementary Fig. 1(d). Below shock pressures of 90 GPa [49], a two-wave elastic-plastic structure is also observed during propagation through the sapphire window (e.g., Supplementary Fig. 9) which is used to determine the shock pressure in the sapphire layer. Further details of pressure calculation is described in Supplementary Fig. 1-2, 5, 7-9.

### D.  First-principles calculations at high pressures

Fe $K$-edge XAS spectra are calculated for the phases listed in Table I. In order to simulate the atomic arrangement in a dense $Fe_{0.6}Mg_{0.4}CO_3$ liquid, first-principles molecular dynamics simulations based on spin-polarized finite-temperature density functional theory (DFT-MD) were performed using the code VASP [50–53].    A 360-atom orthorhombic supercell was constructed from the rhombohedral unit cell of magnesiosiderite at ambient conditions. The exact stoichiometry is 43 Fe atoms, 29 Mg atoms, 72 C atoms and 216 O atoms. We employed the exchange-correlation function of Perdew et al. [54] and modeled the on-site Coulomb repulsion of Fe-localized $d$ atoms within the DFT+$U$ formalism of Duradev et al. [55] with an effective Hubbard correction of +2 eV. We used the PAW pseudopotentials [56, 57] provided with VASP with 8, 8, 4, and 6 valence electrons for Fe, Mg, C, and O atoms, respectively, with a plane-wave energy cutoff of 850 eV. All simulations were performed at the Γ point. The molecular dynamics were done at constant volume and temperature conditions using velocity rescaling (isokinetic ensemble) and a time step of 1 fs. The first simulation was run at 300 K and at ambient density (3.573 g cm$^{-3}$). Then, in order to produce the dense liquid, the initial magnesiosiderite structure was melted at 6000 K and 5.617 g cm$^{-3}$. After this, the volume was adjusted in order to reach a density of 6.234 g cm$^{-3}$ (209 GPa) and the temperature was lowered down to 4000 K, and the system was equilibrated for several picoseconds at these conditions, which are close to validate the following Hugoniot condition [46]:

$$E - E_0 = \tfrac{1}{2}(P + P_0)(\tfrac{1}{\rho_0} - \tfrac{1}{\rho}) \qquad (3)$$

where $E$, $P$, and $\rho$ are the internal energy, pressure, and density, respectively, and subscript 0 refers to the unshocked state at 0 GPa and 300 K.

For each phase of Table I, the isotropic X-ray absorption cross section at the Fe $K$-edge was calculated using the FDMNES code [59, 60] in the finite-difference method approach. Clusters centered on Fe atoms were used, with a cluster radius set to 6 Å, except for $FeCO_3$ at ambient pressure where it was set to 6.65 Å. The electronic potential was calculated using the self-consistent field (SCF) procedure as implemented in FDMNES. The high-spin antiferromagnetic structure of siderite ($FeCO_3$, ambient pressure) was taken into account according to the description given in Badaut et al. [61]. For the other phases, no magnetic structure was introduced in the calculation. For $FeCO_3$, $Fe_4C_3O_{12}$ and $Fe_4C_4O_{13}$, the absorption cross-section was performed in the electric quadrupole approximation, including both $1s \rightarrow 2p$ electric dipole (E1) and $1s \rightarrow 3d$ electric quadrupole (E2) transitions. E2 transitions only contribute to the pre-edge region and were not considered in the spectra calculated for the DFT-MD models. For the DFT-MD structural models (at 300 K and 4000 K), individual absorption spectra of each Fe absorber were first calculated for a few time steps well separated in time (Supplementary Fig. 10). The individual spectra were core-energy shifted with respect to their respective $1s$ level energy, and then averaged. Finally, all the spectra were convoluted using a Lorentzian function with the energy-dependent FWHM:

$$\Gamma(e) = \Gamma_{\text{hole}} + \frac{\Gamma_{\max}}{2} + \frac{\Gamma_{\max}}{\pi} \arctan\left[\frac{\pi}{3}\frac{\Gamma_{\max}}{E_{\text{w}}}\left(e - \frac{1}{e^2}\right)\right] \quad (4)$$

where $e = (E - E_{\text{Fermi}})/E_c$ and with $\Gamma_{\text{hole}}$ = 1.25 eV, $\Gamma_{\max}$ = 10 eV, $E_{\text{w}}$ = 15 eV and $E_c$ = 25 eV.

TABLE I. Phases used for the X-ray absorption cross-section calculations.

| Pressure (GPa) | Chemical composition | Structure used in the XAS calculations |
|---|---|---|
| 0 | $FeCO_3$ | $R\bar{3}c$, 300 K, XRD [58] |
| 0 | $Fe_{0.6}Mg_{0.4}CO_3$ | crystal, 300 K, DFT-MD (this work) |
| 55 | $FeCO_3$ | $R\bar{3}c$, 300 K, XRD [27] |
| 74 | $Fe_4C_3O_{12}$ | $C2/c$, temp.-quenched from 1750(100) K, XRD [22] |
| 97 | $Fe_4C_4O_{13}$ | $R3/c$, temp.-quenched from 1750(100) K, XRD [22] |
| 209 | $Fe_{0.6}Mg_{0.4}CO_3$ | liquid, 4000 K, DFT-MD (this work) |

### E. Recovered samples

Multi-millimeter-thick magnesiosiderite crystals taken from the same rock as in the main experiment at ESRF were embedded in wax inside aluminium holders (Supplementary Fig. 11). The surface on the incoming laser side was polished, and a 70 µm thick layer of Kapton was glued which acted as the ablator. The samples were then shot at the nano2000 facility of the LULI Laboratory in Palaiseau, France, with the spot diameter and laser energy adjusted to match the scaled laser intensities in the main experiment at ESRF (up to $3 \times 10^{12}$ W cm$^{-2}$). Samples were successfully recovered after the shots, the Kapton layer was partially ablated by the laser pulse and ejected from the loaded spot, and a crater was formed inside the sample (e.g., Supplementary Fig. 3). Mapping of the crater floor was performed using Raman spectroscopy.

## III. RESULTS

### A. X-ray absorption spectroscopy under shock loading

Fig. 2(a) shows the normalized XAS spectra and the corresponding approximated pressures of shock-loaded magnesiosiderite (Fe$_{0.6}$Mg$_{0.4}$CO$_3$) with the sample configuration of Fig. 1(a) without the sapphire window. The blue spectra are the reference spectra of unshocked magnesiosiderite taken before the shot and are the average of 10 single X-ray pulse acquisitions. The red spectra are single pulse acquisitions of the corresponding shock-loaded crystal probed just before the shock breakout time inferred from the corresponding Velocity Interferometer System for Any Reflector (VISAR) record, given the delay between the drive laser and the X-ray probe (Supplementary Fig. 1(a) and 1(c)). Each spectrum, except for the highest pressure spectrum, is characterized by three peaks, namely '1' (at ~7127 eV), '2' (at ~7138 eV), and '4' (at ~7175 eV), as shown in Fig. 2(a). The changes in the collected just before (i.e., 0.1 ns to 0.4 ns prior to) shock breakout. The error in timing measurements is ±0.3 ns. Approximated pressures are indicated for the corresponding shock-loaded spectra. **(b)** X-ray absorption spectra of re-shocked magnesiosiderite when using an additional rear layer of sapphire, re-shocked from the shock reverberation at the magnesiosiderite - sapphire interface. The approximated re-shocked pressures are indicated, and positive times indicate that the spectra were collected at time t after shock breakout from the magnesiosiderite layer, or in other words, at time t after the beginning of shock reverberation. **(c)** Calculated X-ray absorption spectra of the crystalline phases at 0, 55, 74, and 97 GPa, and the dense liquid at 209 GPa. Structural information used for calculating the X-ray absorption spectra are listed in Table I. **(d)** Configuration of the dense liquid at 209 GPa and 4000 K, obtained using density functional theory-based molecular dynamics simulation.

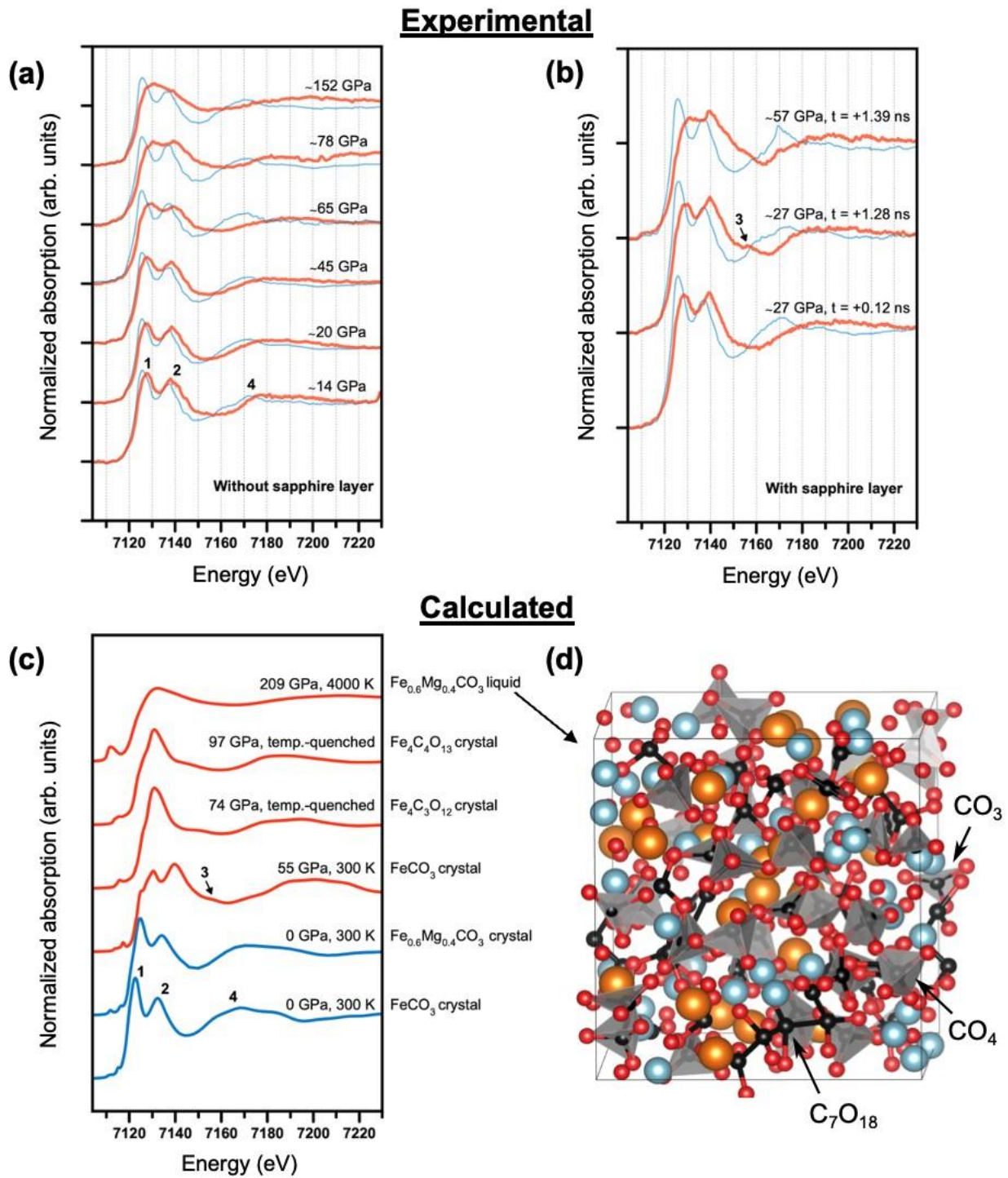

FIG. 2. Experimental and calculated X-ray absorption spectra at ambient and high pressures. **(a)** Pressure evolution of the X-ray absorption spectra of magnesiosiderite under shock loading, without the additional sapphire layer. Blue spectra are the average of 10 × 100 ps single X-ray pulse acquisitions of the magnesiosiderite crystal collected before the shot. Red spectra are the single-pulse acquisitions, 1 x 100 ps, of the shock-loaded spectra of the corresponding magnesiosiderite crystal,

shock-loaded spectra relative to the reference spectra are clearly visible. With increasing pressure: (i) the relative intensities of peaks '1' and '2' decrease, and at the highest pressure, only one peak remains; (ii) the positions of peaks '1' and '2' shift to higher energies; (iii) the peak '4' shifts to higher energy due to increasing density and shortening of the interatomic distances [18]; and (iv) the overall spectrum progressively flattens as a consequence of increasing temperature.

The duration of pressure application is increased when a sapphire window is added to the rear surface of the magnesiosiderite crystal (Fig. 1(a)). This possibly enables the observation of mechanisms that require a longer time under pressure. Additionally, the increase of pressure due to shock reverberation from the magnesiosiderite - sapphire interface also needs to be taken into account. Fig. 2(b) shows the XAS spectra of shots with the additional sapphire layer, where the pressure in magnesiosiderite is calculated using impedance matching as described in section II-C. In the main edge region between 7120 and 7140 eV: (i) the intensities of peaks '1' and '2' become almost equal, even only 0.12 ns after shock breakout from the magnesiosiderite layer and the beginning of shock reverberation; (ii) at the same pressure of ~27 GPa, the intensity of peak '2' becomes higher than that of peak '1' when probed at later times during shock reverberation; (iii) a new peak '3' appears at ~7156 eV at later times during shock reverberation, and (iv) the intensity of peak '2' remains higher than that of peak '1' when probed at similar times, ~1.3 ns, into shock reverberation, but at higher pressure.

### B. Calculated spectra at high pressures

Fig. 2(c) shows the calculated spectra at ambient and high pressures, calculated using the crystalline and dense liquid phases listed in Table I. Similar to the experimental ambient spectra, the calculated ambient spectra are characterized by peaks '1', '2', and '4' at ~7125, 7134, and 7170 eV, respectively. The spectrum at 55 GPa is calculated using the high-pressure $FeCO_3$ structure of Cerantola et al. [27], which represents the electronic high-spin to low-spin transition under static compression. This spectrum shows a higher intensity of peak '2' than peak '1', and the appearance of a new peak '3' at ~7157 eV. Additionally, all the peaks shift to higher energies relative to their positions in the ambient spectra. The spectra at 74 and 97 GPa are calculated using the high-pressure temperature-quenched (from 1750(100) K) structure of Cerantola et al. [22], which represent the structural phase transition of $FeCO_3$ to $Fe_4C_3O_{12}$ and $Fe_4C_4O_{13}$, respectively, both phases containing the $CO_4$ tetrahedra. These spectra are characterized by a single sharper peak in the main edge region between 7120 and 7140 eV. It should be noted that these calculated spectra do not take into account any thermal effects.

The spectrum at the highest pressure of 209 GPa and 4000 K is calculated using our $Fe_{0.6}Mg_{0.4}CO_3$ crystal structure and represents the spectra of a dense $Fe_{0.6}Mg_{0.4}CO_3$ liquid, simulated by DFT-MD simulations as described in section II-D. This spectrum is characterized by a single broad peak in the main edge region, a feature that has been attributed to partial or complete melting under static compression [27]. To assess whether shock-loaded magnesiosiderite is melted at high pressures, we compare our experimental spectrum with the calculated Fe $K$-edge absorption spectrum of the dense $Fe_{0.6}Mg_{0.4}CO_3$ liquid, as shown in Supplementary Fig. 2. The features in the experimental ambient and shock-compressed spectra are reproduced in the calculated spectra. In the main edge region between 7120 and 7140 eV, the double-peak feature in the ambient spectra and the single-peak feature in the shock-compressed spectra are consistent in both the experimental and calculated spectra, along with a general overlap of the absorption edge. In the simulated dense liquid, the total magnetization indicates that Fe atoms are in a low-spin state. Visualization of the atom trajectories reveals that about half of the C atoms adopt a tetrahedral configuration and bond with both O and C atoms forming transient $C_xO_y$ oxycarbon complexes (Fig. 2(d)). It is worth mentioning that the calculated pre-edge in the highest pressure shot (at ~7112 eV) in Fig. 2(c) is likely not relevant, as for the XAS spectra calculations, all Fe atoms of the liquid are considered to have 6 $d$-electrons arranged in a low-spin configuration. However, analysis of the charge density from density functional theory simulations shows that Fe atoms adopt various local charge and spin states depending on the local environment.

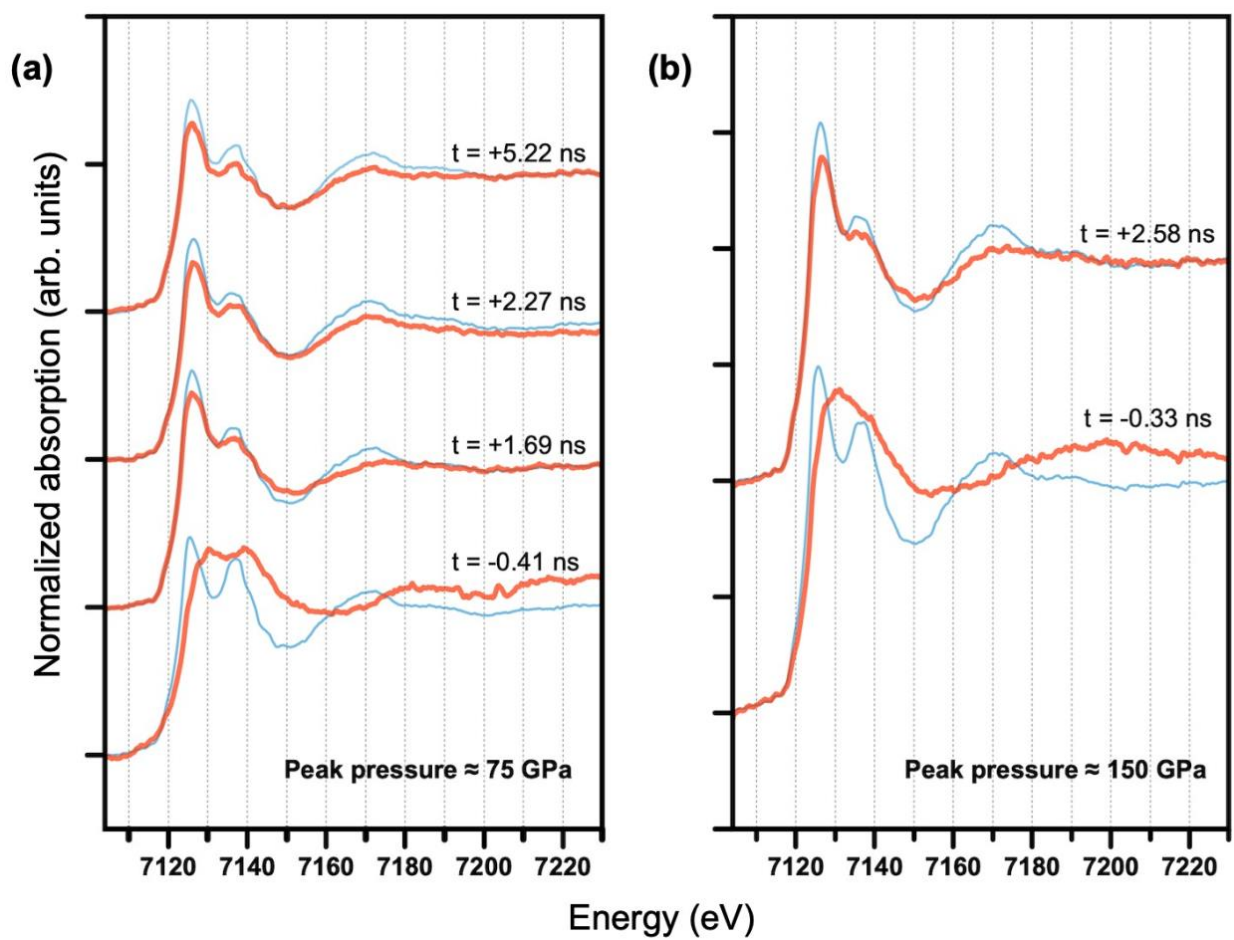

FIG. 3. X-ray absorption spectra during shock unloading. Time scan of the shocked spectra (red) collected at peak pressures of approximately **(a)** 75 GPa and **(b)** 150 GPa. Time values are the delay since the shock breaks out from the magnesiosiderite's free surface: negative when probing the shocked state before breakout, and positive when probing the released states after breakout. Blue

spectra represent the average of the 10 × 100 ps reference spectra of the corresponding unshocked magnesiosiderite crystal.

### C. X-ray absorption spectroscopy during shock unloading

Upon breakout at the free surface, without a sapphire window, the incident compression front is reflected into a release wave, which gradually unloads the sample back to zero pressure. Thus, by increasing the delay between the drive laser and the X-ray probe, the material can be probed during pressure release from the shocked state. Unexpectedly, probing under release conditions shows spectra very close to the ambient magnesiosiderite spectra. Fig. 3 shows the time scan of the XAS spectra collected at two different laser intensities corresponding to peak shock pressures of approximately 75 GPa and 150 GPa. Time values indicated in Fig. 3 are the delay since the shock breaks out from the magnesiosiderite layer: negative when probing the shocked state before breakout, and positive when probing the released states after breakout. Independent of the peak pressure, moving from negative time (shock loading) to positive time (shock release) causes the two peaks, '1' and '2', to shift back to lower energies at positions corresponding to those of the ambient spectra shown in blue. Additionally, the intensity of peak '1' becomes higher than that of peak '2', matching the scheme of the ambient spectra. Since entropy intrinsically jumps across a shock front and the release wave follows a quasi-isentrope, the temperature under release is significantly higher than the ambient temperature. This explains the spectral features being less prominent in the shocked spectra compared to the corresponding reference spectra due to the Debye-Waller effect [62].

### D. Raman spectroscopy on recovered samples

A few multi-millimeter-thick magnesiosiderite crystals taken from the same rock were shock-compressed at the nano2000 facility of the LULI Laboratory. Laser energy and spot diameter were adjusted to match the same scaled intensities (up to $3 \times 10^{12}$ W cm$^{-2}$) as in the XAS results collected in the main experiment at ESRF. Samples were recovered after shock compression to perform post-shock analyses. The recovered samples show a crater on the loaded surface with dimensions similar to the expected size of the irradiated spot (Supplementary Fig. 3). The fracture surface morphology at the bottom of the craters shows an anisotropic brittle behavior with straight parallel cracks in specific directions related to the crystal orientation. No obvious evidence of melting or resolidification was observed inside these cracks or between them. The formation of these craters results from (i) the motion of the shock-compressed material, and (ii) shock-induced fragmentation commonly observed in brittle materials (likely enhanced by the contribution of lateral release waves at late times) followed by possible loss of some fragments upon ejection of the Kapton (ablator) foil. If such loss is disregarded, the material at the bottom of the crater is expected to have experienced a thermodynamic history (shock compression and release) roughly similar to that in the main experiment with similar laser conditions.

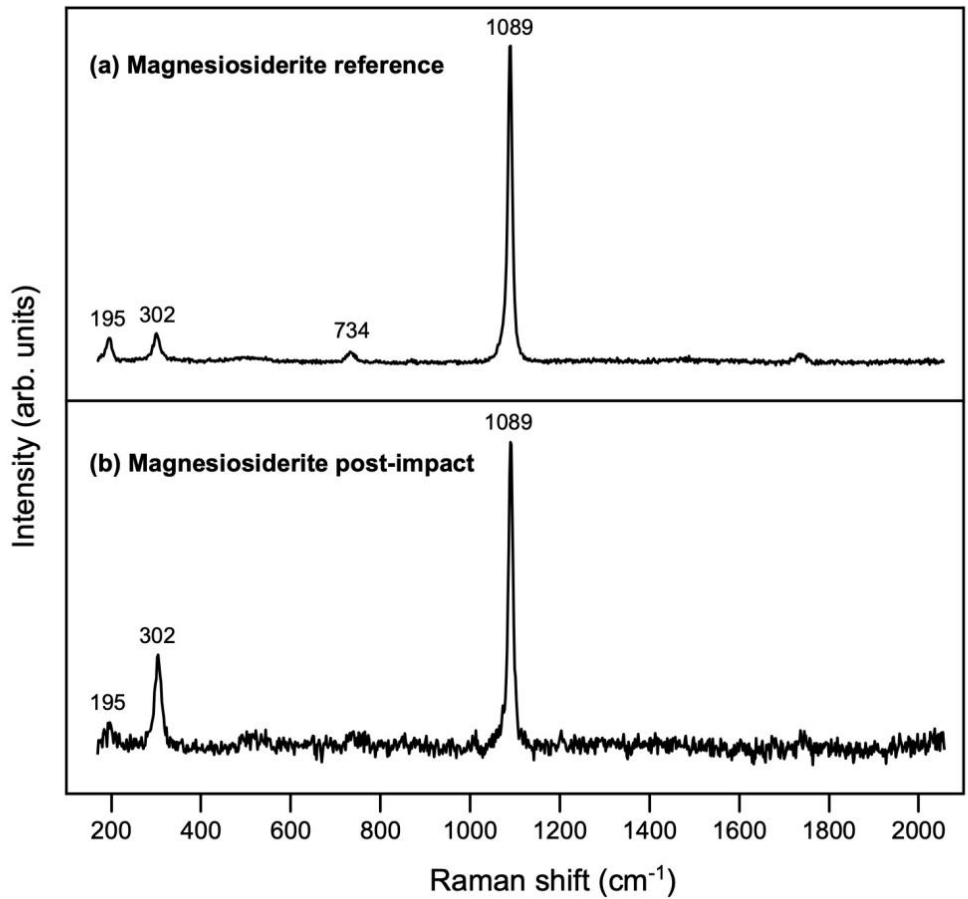

FIG. 4. Raman spectrum of post-impact recovered magnesiosiderite. (a) Reference Raman spectrum of the unshocked magnesiosiderite crystal. The bands at 195, 302, 734 and 1089 cm$^{-1}$ coincide with previously reported bands of natural siderite [63, 64]. **(b)** Raman spectrum of the crater floor in the corresponding post-impact recovered magnesiosiderite crystal, shock compressed using a laser intensity of $1.1 \times 10^{12}$ W cm$^{-2}$, corresponding to a pressure of ~20 GPa.

Extensive mapping of the crater floor in the post-impact recovered samples was carried out using Raman spectroscopy. No spectrum significantly different from the reference magnesiosiderite was detected (e.g., Fig. 4), apart from a few rare hematites, $Fe_2O_3$ (Supplementary Fig. 4). The Mg content in the recovered sample was similar to that in the initial sample. No detectable magnetite ($Fe_3O_4$) was observed.

## IV. DISCUSSION

### A. Spin transition in shock-loaded magnesiosiderite

Spin crossover in siderite and magnesiosiderite has been extensively studied under static compression using various experimental techniques such as X-ray diffraction [11, 12, 15, 16, 40], X-ray absorption and Mössbauer spectroscopy [18], Raman spectroscopy [15, 18, 24], Xray Raman scattering [26], and X-ray emission spectroscopy [28, 65]. This pressure-induced high-spin to low-spin transition, associated with a ~8-10% volume collapse [11, 15, 29], results from the electronic spin crossover of $3d$ electrons of $Fe^{2+}$ atoms [18]. The compression of the $FeO_6$ octahedra with increasing pressure causes an increase in the crystal field splitting energy between the triply degenerate $t_{2g}$ and doubly degenerate $e_g$ levels within the $3d$ orbitals. This leads to a rearrangement of the $3d$ electrons from a $t_{2g}^4 e_g^2$ configuration with four unpaired electrons (high-spin) to a $t_{2g}^6 e_g^0$ configuration with no unpaired electrons (low-spin).

Under static compression, spin crossover in siderite occurs at ~45 GPa at room temperature [11, 15, 18]. This pressure shifts to ~53 GPa at elevated temperatures of ~1200 K, accompanied by a broadening of the pressure range for the spin crossover [16]. In the FeCO$_3$-MgCO$_3$ solid solution, there is no observable compositional effect on the spin transition pressure at room temperature [10, 15, 16]. Additionally, the spin transition remains independent of temperature up to 700 K [24].

Under dynamic compression, the first experimental Hugoniot data of an iron-rich carbonate, (Fe$_{0.75}$Mg$_{0.25}$)CO$_3$, has recently been reported by Wang et al. [31] using a two-stage light gas gun setup. In their shock velocity vs. particle velocity ($U_S$ - $u_p$) data, Wang et al. [31] report a discontinuity at $u_p$≈1.5 km s$^{-1}$, corresponding to a pressure of 38 GPa. By comparison with static compression data, they attribute this discontinuity to the electronic high-spin to low-spin transition. Additionally, Wang et al. [31] report a very sharp spin transition, with a transition range of less than 1.6 GPa, and a calculated shock temperature of approximately 500 K at the transition pressure. This transition is also associated with a ~7% volume change [31].

Our XAS results provide *in situ* X-ray characterization of a shock-compressed iron-rich carbonate and allow for a qualitative analysis of the spin crossover based on the changes in the XAS features from static compression. With increasing pressure under shock compression (Fig. 2(a)), the shifts of peaks '1' and '2' to higher energies, along with the increasing intensity of peak '2' relative to peak '1', match the trends observed in the XAS spectra under static compression [18]. These changes are associated with the unit-cell compression and the resulting decrease in the Fe-O bond length, which are related to the electronic transition from the high-spin state to the low-spin state [11, 15, 18, 29].

Under static compression, peak inversion occurs where the intensity of peak '2' becomes higher than that of peak '1', and a new peak '3' appears at ~7157 eV at a pressure of 44 GPa and room temperature [18]. This is identified as a signature of structural changes, primarily a volume collapse associated with the spin transition, complemented by Raman and Mössbauer spectroscopy [18]. Peak '3' could arise from multiple scattering of the photoelectrons from neighboring oxygen atoms, which move closer to the iron atoms after the spin transition [18].

Under shock compression, while the relative intensity of peak '2' increases with increasing shock pressure (Fig. 2(a)), it does not surpass the intensity of peak '1', and the new peak '3' does not appear. This suggests that the spin crossover is not complete even at higher pressures, possibly due to ultrafast pressure release occurring after the shock breakout.

While the completion of the high-spin to low-spin transition is not observed under shock loading, further low-spin nucleation is observed under shock reverberation when an additional rear layer of sapphire is used. In Fig. 2(b), when staying at the same pressure of ~27 GPa but probing at later times during shock reverberation, the exact features appear in the XAS spectra that have been attributed to the spin transition under static compression. Specifically, the intensity of peak '2' becomes higher than that of peak '1', and a new peak '3' appears at ~7157 eV. Although precise pressures are difficult to determine, our measurements, with an uncertainty of ±15 GPa (Supplementary Fig. 5), fall within the range of spin transition pressures observed under static compression [10, 12, 15, 16, 18, 26] and suggested under dynamic compression [31]. These spectral features persist at higher re-shock pressures (Fig. 2(b)) but are suppressed at similar pressures without the re-shock effect (Fig. 2(a)).

The differences in the spectral features with and without shock reverberation could have various origins. The simplest explanation is the difference in the pressure-temperature paths between single shock loading and shock reverberation. At a given pressure, the re-shocked state is denser and at lower temperatures than the single-shocked state. For example, the middle spectra in Fig. 2(b) transitions from a single-shocked pressure of ~24 GPa, at t = 0 ns, to a re-shocked pressure of ~27 GPa after 1.28 ns. The temperature at 24 GPa under single-shock loading is ~410 K, estimated from the calculated quotidian equation of state (QEOS) of siderite (Supplementary Fig. 6).

Since the pressure jumps due to shock reverberation are not large, and the spectral features attributed to spin transition are not observed even at higher pressures under single shock loading (e.g., 78 GPa spectra in Fig. 2(a)), an alternate explanation is that the spin transition might be a kinetics-dependent process. The high pressure and temperature conditions maintained for a longer time when using the additional rear sapphire window, lasting a few nanoseconds compared to a few hundred picoseconds, could be a prerequisite for the spin transition. The argument for kinetics is further supported by the inference of the spin transition using a two-stage light gas gun setup [31], where shock compression occurs over microsecond-order timescale compared to the nanosecond timescale in our laser-driven shock compression experiment. Further experiments are needed to distinguish the kinetic and thermodynamic dependence of the spin transition in iron-bearing carbonates.

### B. Structural phase transition in shock-loaded magnesiosiderite

Under static compression, the spin transition in siderite and magnesiosiderite is also associated with structural phase transitions at high pressures and temperatures. At approximately 50 GPa and 1400 K, a structural transition from the rhombohedral phase-I (space group $R\bar{3}c$) to the orthorhombic phase-II was reported in siderite, possibly driven by the spin transition of iron [40]. In the same study, magnesiosiderite-II with a composition $Fe_{0.65}Mg_{0.35}CO_3$ was detected starting from 60 GPa and 1500 K. The orthorhombic phase-II of siderite [40] was later confirmed to be iron tetracarbonate (space group $R\bar{3}c$), containing $CO_4$ tetrahedra [22]. This phase containing the $CO_4$ tetrahedra is also observed in natural magnesiosiderite at pressures above 80 GPa and temperatures above 1850 K [30].

Under dynamic compression of $(Fe_{0.75}Mg_{0.25})CO_3$, Wang et al. [31] report a second discontinuity in their Hugoniot data at $u_p \approx 2.4$ km s$^{-1}$, corresponding to a shock pressure of 65-69 GPa. With a calculated temperature of 1270±190 K, they attribute this discontinuity at 65 GPa to a self-redox reaction. Furthermore, Wang et al. [31] report that the volume change during this self-redox transition are consistent with the reaction products of tetrairon orthocarbonate, $Fe_4C_3O_{12}$, and diamond.

However, during our shock compression experiment on $Fe_{0.6}Mg_{0.4}CO_3$, the smooth and continuous changes observed in our spectra (Fig. 2(a)) up to ~78 GPa suggest no structural phase transition. Instead, our molecular dynamics simulations indicate the transition to a $CO_4$ tetrahedra-containing phase in the shock-induced melt (Fig. 2(d)). At 74 GPa, although our calculated spectra (Fig. 2(c)) is constructed using the temperature-quenched high-pressure phase of $Fe_4C_3O_{12}$ of Cerantola et al. [22], its spectral features are distinctly different from the experimental spectra at a similar pressure of 78 GPa (Fig. 2(a)). This further suggests that the transition to the teracarbonate phase does not occur at this pressure in the crystalline phase under nanosecond single-shock loading.

While Wang et al. [31] suggest a phase transition to tetracarbonates in the crystalline phase at a pressure of ~65 GPa, our nanosecond compression indicates that this transition occurs in the shock-induced melt at pressures above 100 GPa (Fig. 2(c,d)). Additionally, in the similar pressure range of ~60 GPa, the tetracarbonate phase appears at around 1500 K under static compression [40], but at around 1270 K under sub-microsecond shock compression [31]. These differences can be attributed to different pressure-temperature paths, strain rates, phase transition mechanisms, Mg content, or reaction kinetics. In particular, if the transition kinetics are slow enough, i.e., slower than the shock wave propagation (a few nanoseconds), it is likely that the transition to the 'high-pressure' phase does not occur or might happen directly in the melt, provided no dissociation reactions take place. Future shock compression experiments coupled with *in situ* X-ray diffraction will be helpful in resolving uncertainties related to shock-induced structural transitions.

### C. Shock-induced melting and transformation in magnesiosiderite

Under static compression, siderite melting is not stoichiometric but partially incongruent, and the presence of minor quenched magnetite has always been observed [19]. This phenomenon is attributed to the partial redox dissociation of $FeCO_3$-melt, resulting in dissolved $Fe^{3+}$ and $CO_2$ in the carbonate melt, with less than 10% ferric iron components [19]. Even after prolonged heating in static compression, the decomposition to Fe-oxides is never complete, and upon quenching, recrystallized $FeCO_3$ and $Fe_2O_3$, or high-pressure $Fe_3O_4$ are always observed [19, 22, 27, 66]. Recovered $Fe_3O_4$ and $Fe_2O_3$ are also found after heating and melting siderite at 51 GPa [28].

Under shock compression, the features in the XAS spectra, shock-compressed to the highest pressure of ~150 GPa in Fig. 2(a), resemble those observed under static compression in a laser-heated diamond anvil cell, which were previously attributed to complete or at least partial melting [27]. Our molecular dynamics simulations, which modeled a dense $Fe_{0.6}Mg_{0.4}CO_3$ liquid, suggest that in the melt, about half of the C atoms adopt a tetrahedral configuration. During the transformation to tetracarbonates in the melt, the coordination of carbon increases, facilitated by the ability of $CO_4$ to form polymerized networks, unlike the $CO_3$ trigonal groups [17, 67, 68]. The change in carbon atom hybridization from $sp^2$ to $sp^3$, necessary for the transition to tetracarbonates, was observed in crystalline samples and also during the compression of carbonatitic glass $K_2Mg(CO_3)_2$ above 40 GPa, used as an analog for carbonate melts [68].

In the two previous shock recovery experiments on natural siderite, magnetite ($Fe_3O_4$) or magnetite-like phases were detected in the recovered samples [32, 33]. In the nanosecond laser-driven shock experiment [32], the starting siderite sample contained hematite ($Fe_2O_3$), suggesting that the magnetite-like phase in the recovered samples may be a redox product of hematite in the starting material [66]. The impact-driven projectile experiment by Bell [33] was performed in a $CO_2$ atmosphere ($10^{-3}$ torr), where the starting sample was free of detectable magnetite. The transformation of siderite to magnetite was detected in the recovered sample after compression to 49 GPa, likely due to local heating above 743 K [33].

In our nanosecond laser-driven shock experiment, with the starting samples also free of detectable iron oxides, no decomposition to magnetite is observed, either under shock loading or the pressure release, or in the recovered samples. Instead, under pressure release, we observe signatures of magnesiosiderite recrystallization (Fig. 3). Our results are also in contrast to the flyer plate impact experiment on magnesite ($MgCO_3$), where MgO crystallites were detected in the recovered samples shocked to above 95 GPa, with the amount of MgO increasing with increasing shock pressure [69]. While the low siderite-to-magnetite decomposition pressures observed by Bell [33] could be attributed to the use of powdered samples, resulting in higher shock temperatures [31], it is also possible that the decomposition to magnetite occurs over longer timescales or during temperature release.

## V. CONCLUSION

In this study, we have investigated the properties of magnesiosiderite, $Fe_{0.6}Mg_{0.4}CO_3$, under nanosecond laser-driven shock compression, coupled with *in situ* X-ray characterization using ultrafast synchrotron X-ray absorption spectroscopy (XAS). Under single shock loading, the XAS spectra suggest that $Fe^{2+}$ in magnesiosiderite undergoes a partial high-spin to low-spin transition. The signatures of the spin-transition become more pronounced under shock reverberation, suggesting thermodynamic (associated with temperature) and/or kinetic effects. The phase of the magnesiosiderite crystal remains unchanged up to the melt. In the shock-induced melt at pressures above 100 GPa, our DFT-MD simulations indicate the formation of iron tetracarbonates containing the $CO_4$ tetrahedra. No detectable decomposition to magnetite ($Fe_3O_4$) was observed, consistent with Raman spectroscopy analysis of the recovered samples. Upon unloading from the shocked state, magnesiosiderite recrystallizes independent of the peak shock pressure. These results provide valuable guidance for future *in situ* investigations of iron-bearing carbonates under varying strain rates and time scales, given their significant geological and extraterrestrial implications.

## ACKNOWLEDGMENTS

The authors would like to thank the ESRF and Gaston Garbarino for providing beamtimes on ID24-ED (HC-4714) and ID15B beamlines. The authors are thankful to Leonid Dubrovinsky for supervising the crystal selection, the machine shop at Bayerisches Geoinstut for polishing the crystals, and the LULI2000 staff for the shots on the recovered samples. A.P.D. acknowledges funding from the European XFEL. J.-A.H. acknowledges access to the ESRF supercomputing resources. D.Ca. and M.H. acknowledge the MESU platform of Sorbonne Universit´e for providing access to supercomputing resources and are thankful to Yves Joly (Institut Néel) for fruitful discussions.

**Author contributions:** R.T., T.d.R., and F.G. initiated the project. H.M. provided and characterized (FT-IR, SEM/EDX, and P-XRD) the magnesiosiderite samples. D.Co. performed the single crystal X-ray diffraction measurements and structural refinement. I.S. and F.G. performed SEM/EDX characterization of the magnesiosiderite crystals. A.P.D selected the crystals under the supervision of V.C. M.H. prepared the samples for the main experiment. J.-A.H., N.S.-R., R.T. and S.B. operated the beamline for the main experiment. A.P.D. J.-A.H., T.d.R., S.B., and M.H. performed data analysis with inputs from V.C., D.Ca., F.G., N.S.-R., and R.T. S.B. and J.-A.H. prepared the samples for the recovery experiment; T.d.R. and T.V. performed the recovery experiment; T.d.R. performed initial post-recovery analysis. F.G. and D.D. performed Raman analysis of the recovered samples. T.V. performed calculations for the QEOS of siderite. J.-A.H. performed DFT-MD simulations. D.Ca., J.-A.H., and M.H. performed first-principles calculations. A.P.D. wrote a draft of the manuscript, and all the co-authors read, revised, and commented on it.

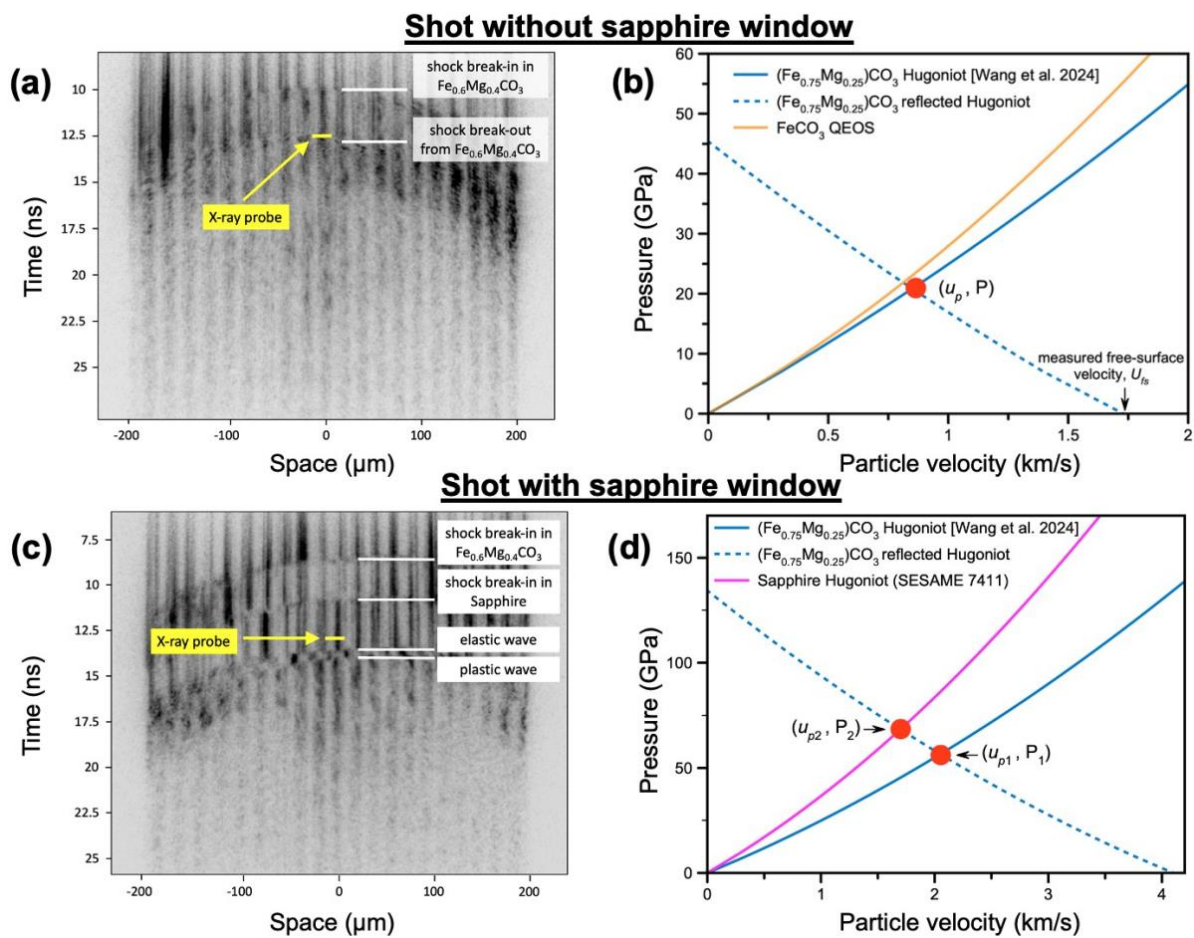

**Supplementary Fig. 1. Pressure determination from VISAR data. (a)** VISAR data and **(b)** pressure determination for the sample configuration without the sapphire layer (configuration of Fig. 1(a) without the sapphire layer). To avoid uncertainties that make their way into the pressure calculation, such as the variation in thickness from one crystal to another and the unknown thickness of the glue layer between the ablator and the crystal, we use the measurement of the free surface velocity ($U_{fs}$) from the VISAR to calculate the pressure. The particle velocity ($u_p$) is calculated as $u_p = U_{fs}/2$ [47, 48], where the free surface velocity ($U_{fs}$) is measured from the VISAR data. For higher pressure shots, and for shots where a clear free surface velocity could not be measured, a linear fit to the $u_p$ vs. effective laser intensity plot (Supplementary Fig. 7) is used to infer the particle velocity $u_p$. This $u_p$ value gives the shock pressure, resulting in the same value of pressure obtained using the method described in section II-C of the main text. For comparison with the experimental Hugoniot data of $Fe_{0.75}Mg_{0.25}CO_3$ of Wang et al. [31], the calculated Quotidian Equation of State (QEOS) of $FeCO_3$ (Supplementary Fig.

6) is shown in orange. **(c)** VISAR data for a shot with the additional layer of sapphire glued on top of the magnesiosiderite crystal. A double-wave structure of elastic and plastic waves is observed at the sapphire-vacuum interface. This double-wave structure is used to determine the shock pressure in the sapphire window, as described in Supplementary Fig. 9. **(d)** Impedance matching construction for pressure determination for the sample configuration with the additional sapphire layer (configuration of Fig. 1(a)). Magnesiosiderite, $Fe_{0.6}Mg_{0.4}CO_3$, has a lower impedance compared to sapphire, as evidenced by the slope of the Hugoniot curves. When the shock wave travels from magnesiosiderite to sapphire, a reverberation of the shock occurs at the magnesiosiderite - sapphire interface, which takes magnesiosiderite from the state ($u_{p1}$, $P_1$), along the reflected Hugoniot curve, to state ($u_{p2}$, $P_2$), matching the shocked state in sapphire. Hence, at times later than the shock breakouts from magnesiosiderite, it will be in the state ($u_{p2}$, $P_2$).

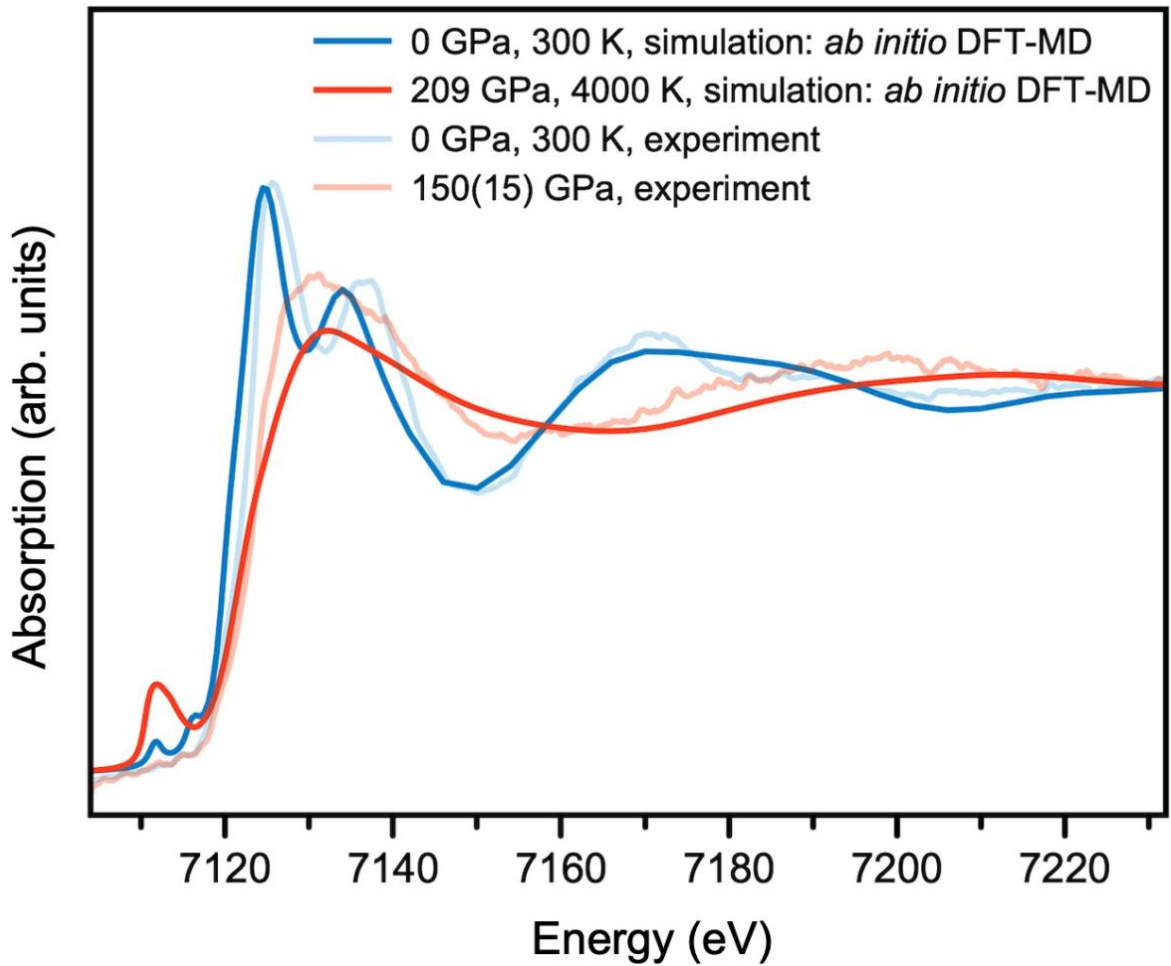

**Supplementary Fig. 2. Comparison between experimental and calculated X-ray absorption spectra at the highest pressure.** DFT-MD calculations to simulate a dense $Fe_{0.6}Mg_{0.4}CO_3$ liquid were

performed at 209 GPa. At the time of performing these simulations, the Hugoniot data of (magnesio)siderite of Wang et al. [31] were not available, therefore, the Hugoniot data of magnesite ($MgCO_3$) [70] was used to approximate pressures, giving a value of ~200 GPa. The pressure value of 150(15) GPa for the experimental data is obtained using the method described in section II-C of the main text and Supplementary Fig. 1. Instead, if the thickness of the magnesiosiderite crystal (~21 μm) and the transit time of the shock wave across this thickness (2.01 ns) is used to calculate the shock velocity (=10.44 km $s^{-1}$), and hence the particle velocity (using equation 2) and pressure (equation 1), a value of ~129 GPa is obtained for the pressure. A general overlap of the spectral features between the calculated and experimental spectra suggests that at the highest pressure of 150(15) GPa, magnesiosiderite was at least partially melted. It should be noted that the calculations are not optimized for the pre-edge region (below 7120 eV).

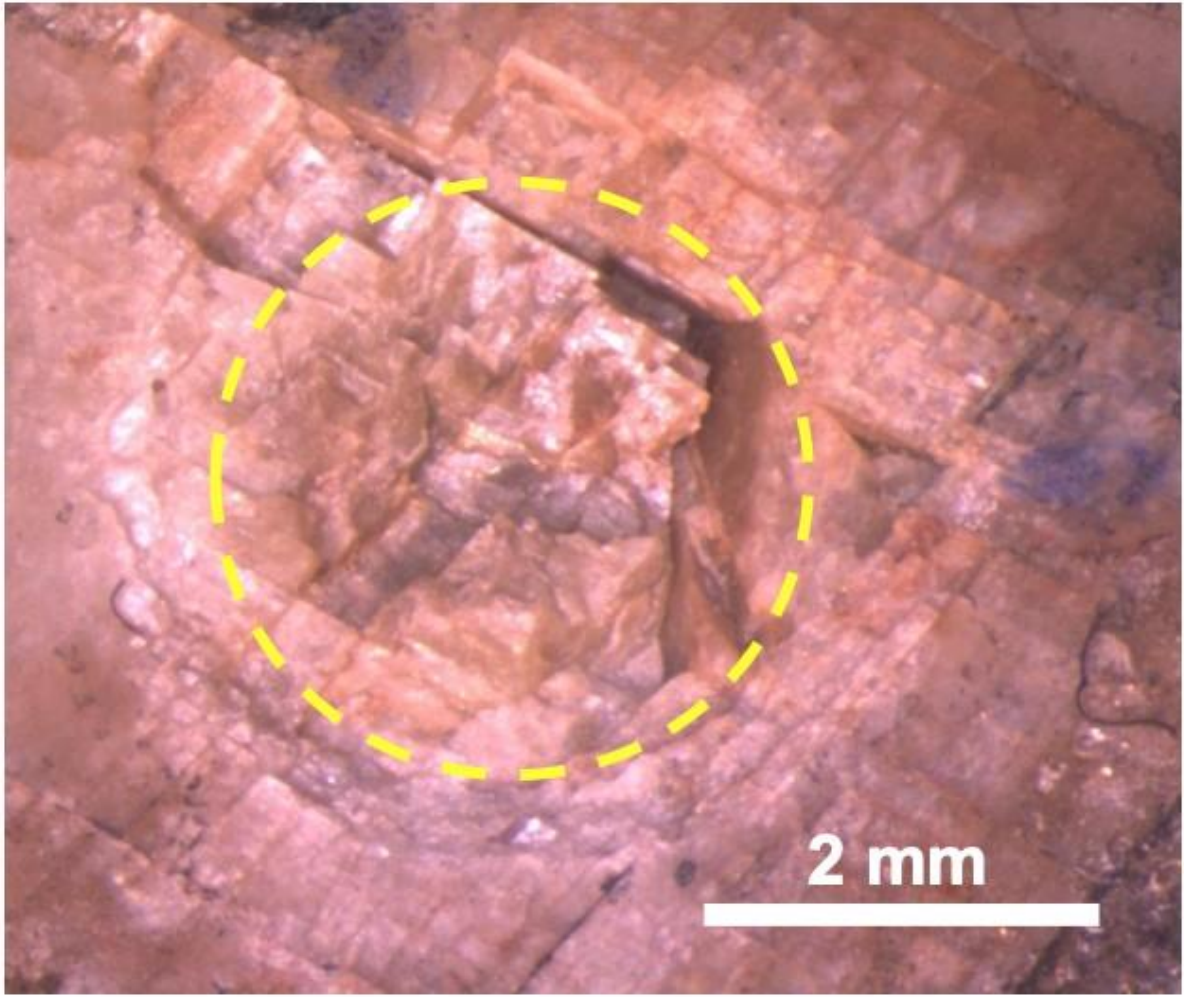

**Supplementary Fig. 3. Crater in the loaded surface in the post-impact recovered sample.** Typical crater observed in the loaded surface of the multi-millimeter thick magnesiosiderite sample recovered after being subjected to a laser shock at the LULI nano2000 facility. Laser intensity for this shot was $1.10 \times 10^{12}$ W cm$^{-2}$. Sample and experimental configuration of the recovery shots is shown in Supplementary Fig. 11. The yellow circle represents the diameter of the laser spot, 3 mm in this case. The Kapton foil used as an ablator is ejected from the shock-loaded region, as expected from the impedance mismatch at the interface with the sample.

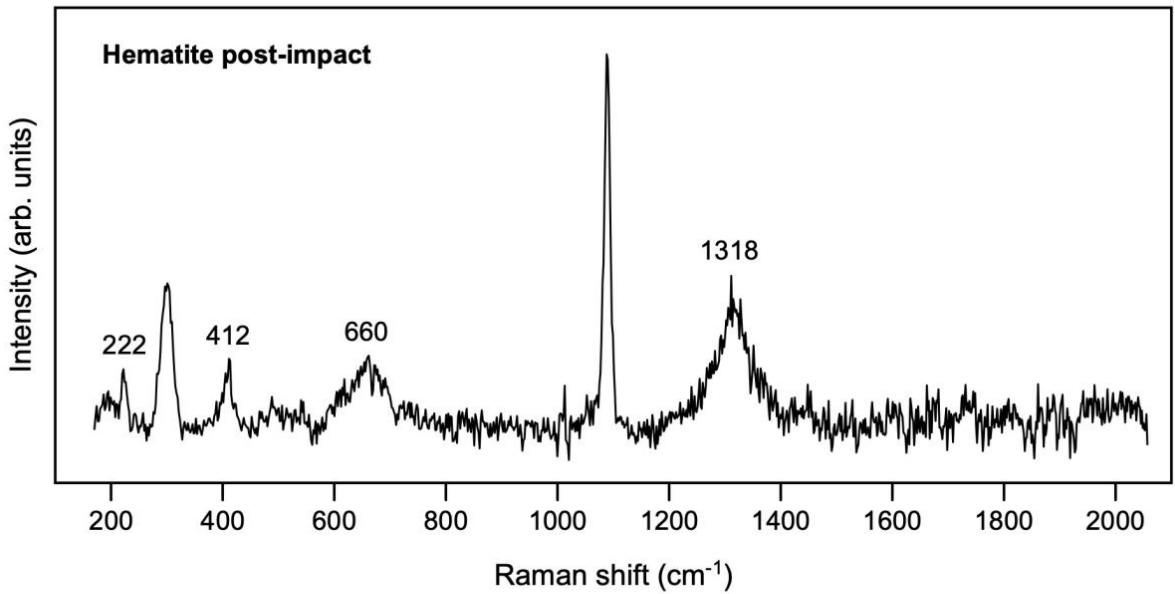

**Supplementary Fig. 4. Raman spectrum of the post-impact recovered magnesiosiderite crystal shows a magnesiosiderite-hematite mix.** Extensive mapping of the crater floor (e.g., Supplementary Fig. 3) in the post-impact recovered magnesiosiderite crystals show a few rare crystallites of hematite, $Fe_2O_3$. The bands at 222, 412, and 1318 cm$^{-1}$ are assigned to hematite [72]. Hematite also has a band at 612 cm$^{-1}$ [72], which is likely encompassed in the broad band at 660 cm$^{-1}$. While magnetite ($Fe_3O_4$) also has a band at ~660 cm$^{-1}$ [63], the appearance of the band at this position in our spectrum cannot be conclusively attributed to magnetite.

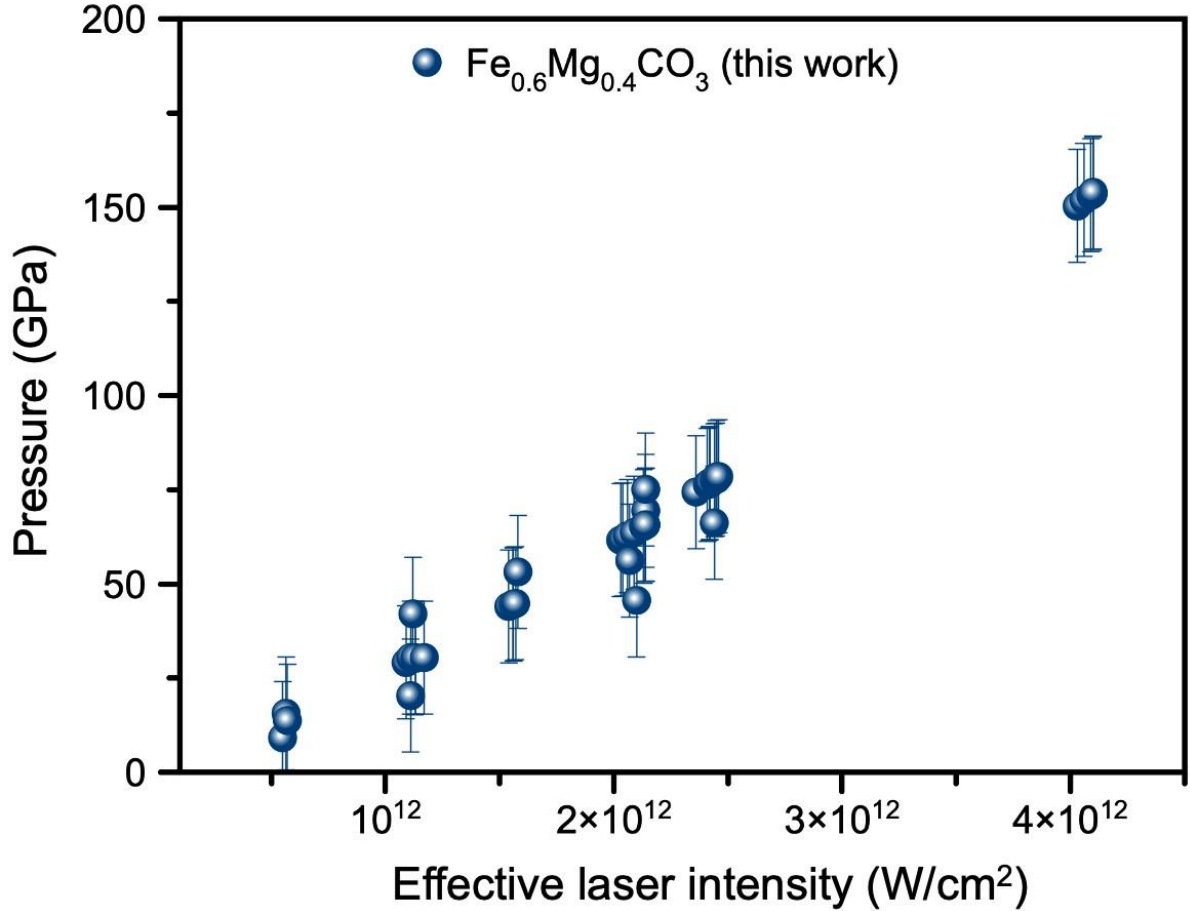

**Supplementary Fig. 5. Error in pressure.** Multiple shots were performed at the same laser intensity. Pressure values were calculated using the VISAR data and the method described in section II-C of the main text. The largest variation in the value of measured particle velocity ($u_p$), and hence the pressure, is used as the error value for all pressure points. This largest variation in $u_p$ occurs at the laser intensity of 2.3 x $10^{12}$ W cm$^{-2}$ (Supplementary Fig. 7), and gives a value of ±15 GPa as the error in pressure.

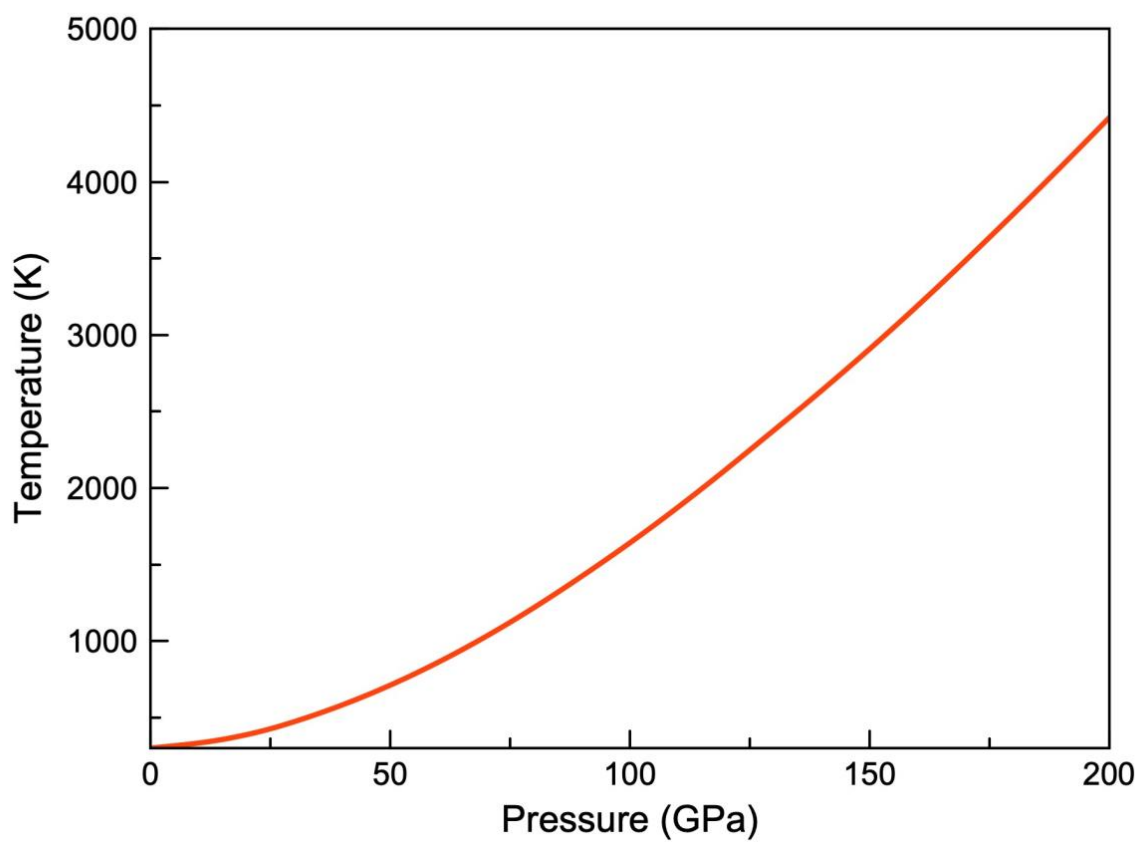

**Supplementary Fig. 6. Pressure vs. temperature curve using the calculated QEOS of FeCO$_3$.** The Quotidian Equation Of State of siderite was constructed using the Thomas Fermi model and the following parameters: effective atomic number=11.6, average molar mass = 23.6, solid density = 3.96 g cm$^{-3}$, bulk modulus = 124 GPa.

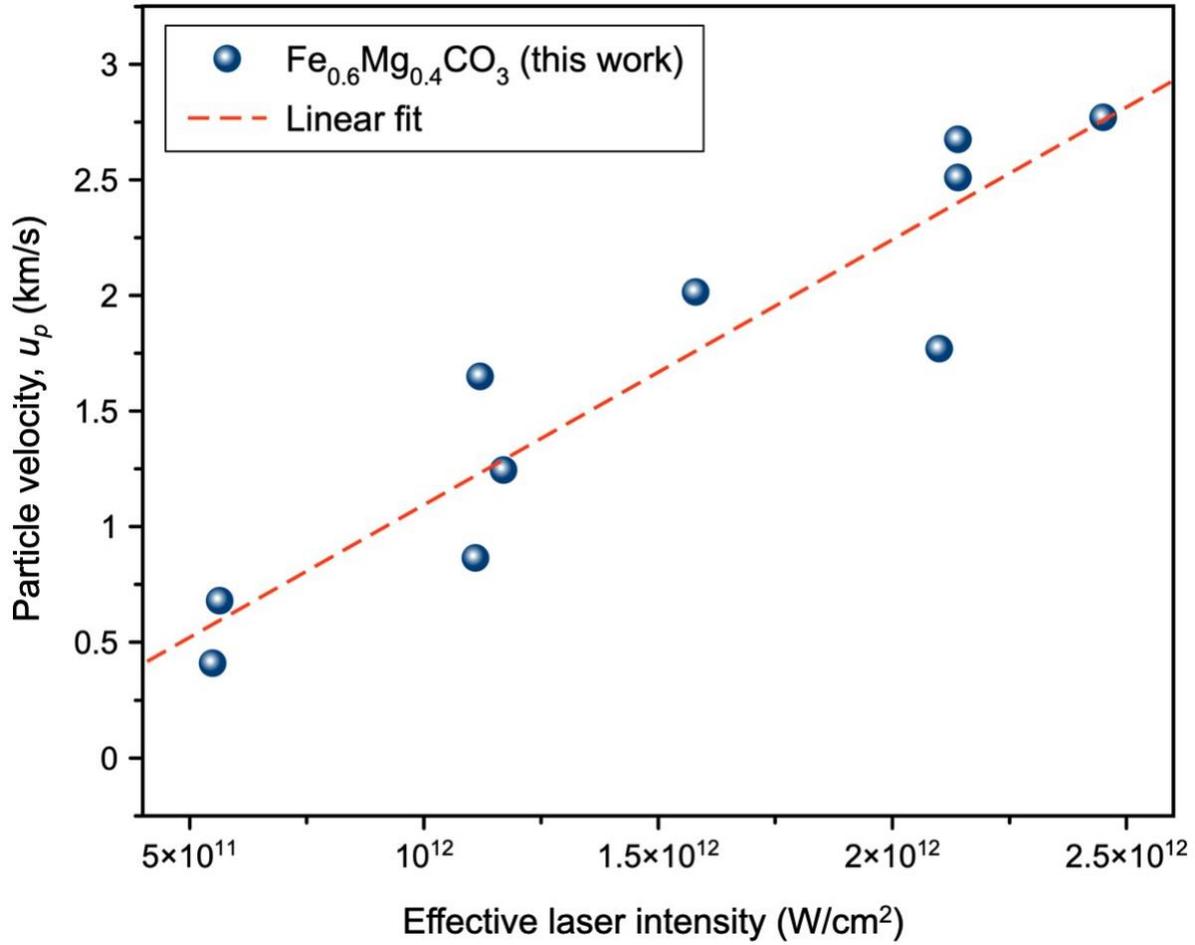

**Supplementary Fig. 7. Fit to the particle velocity ($u_p$) vs. effective laser intensity.** A clear magnesiosiderite free surface velocity ($U_{fs}$) was measured for these shots using the VISAR, and the particle velocity was calculated as $u_p = U_{fs}/2$ [47, 48]. The fit to this plot is used to infer the particle velocity for higher pressure shots and for shots where a clear and reliable free surface velocity could not be measured.

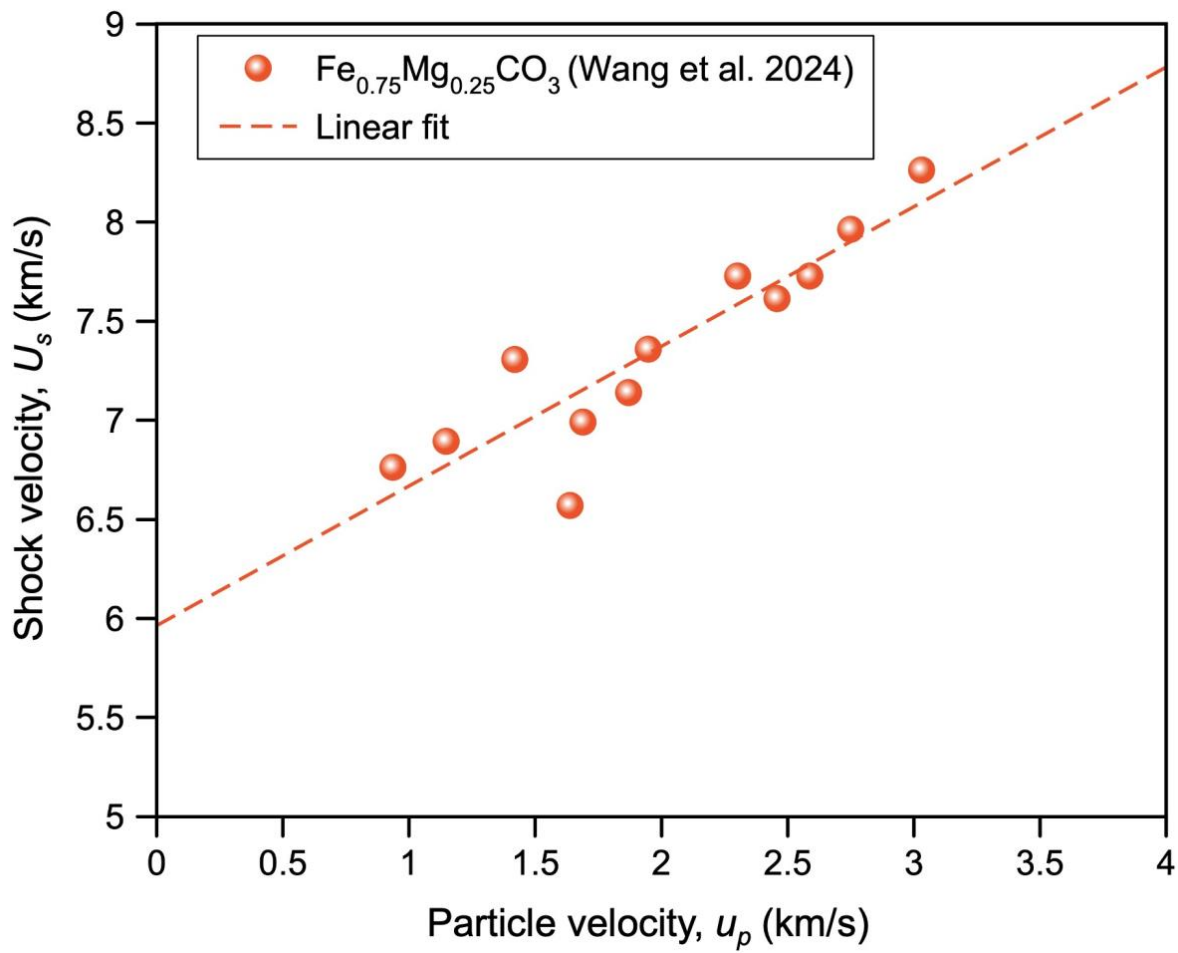

**Supplementary Fig. 8. Linear fit to the $Fe_{0.75}Mg_{0.25}CO_3$ Hugoniot data of Wang et al. [31].** Wang et al. [31] report two discontinuities in their data at $u_p \approx 1.5$ and 2.4 km/s, which they attribute to the spin transition and the self-redox reaction, respectively, and document three piecewise linear fits to their data. However, since our data suggests an incomplete spin-transition under single shock loading, and the self-redox reaction, which Wang et al. [31] link to the formation of tetracarbonates in the crystalline phase, is absent in the crystalline phase of our data, we disregard the three linear piecewise fits but instead consider the entire $U_s$-$u_p$ data range of Wang et al. [31]. All minor pressure differences are within error when considering the fit over the whole data range compared to the three piecewise linear fits, and hence do not result in any changes in the results or their interpretation.

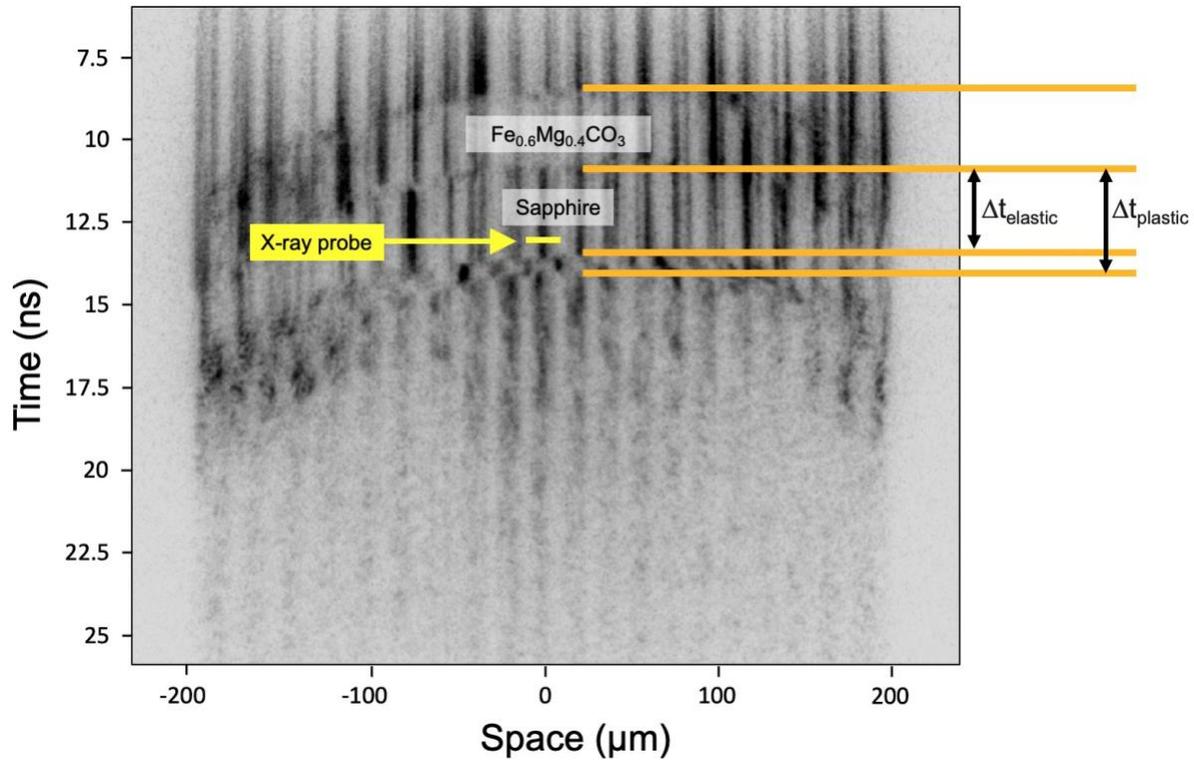

**Supplementary Fig. 9. VISAR record with an additional layer of sapphire on top of the magnesiosiderite crystal.** The shock wave entrance in the magnesiosiderite and sapphire layers is shown. At the sapphire - vacuum interface, a double-wave structure of the elastic and plastic waves is observed, and the transit times of these waves in the sapphire layer are indicated. The transit time of the elastic wave in sapphire, $\Delta t_{elastic}$, and the elastic wave velocity in sapphire of ~11.45 km/s [71] is used to calculate the thickness of the sapphire layer. This value matches well with the measured thickness of the sapphire layer before the experiment, with an uncertainty of ~3%. This calculated thickness of the sapphire layer, and the measured transit time of the plastic wave, $\Delta t_{plastic}$, is used to calculate the plastic wave velocity in sapphire, which is then used in the impedance matching (Supplementary Fig. 1(d)) to determine the pressure in re-shocked magnesiosiderite.

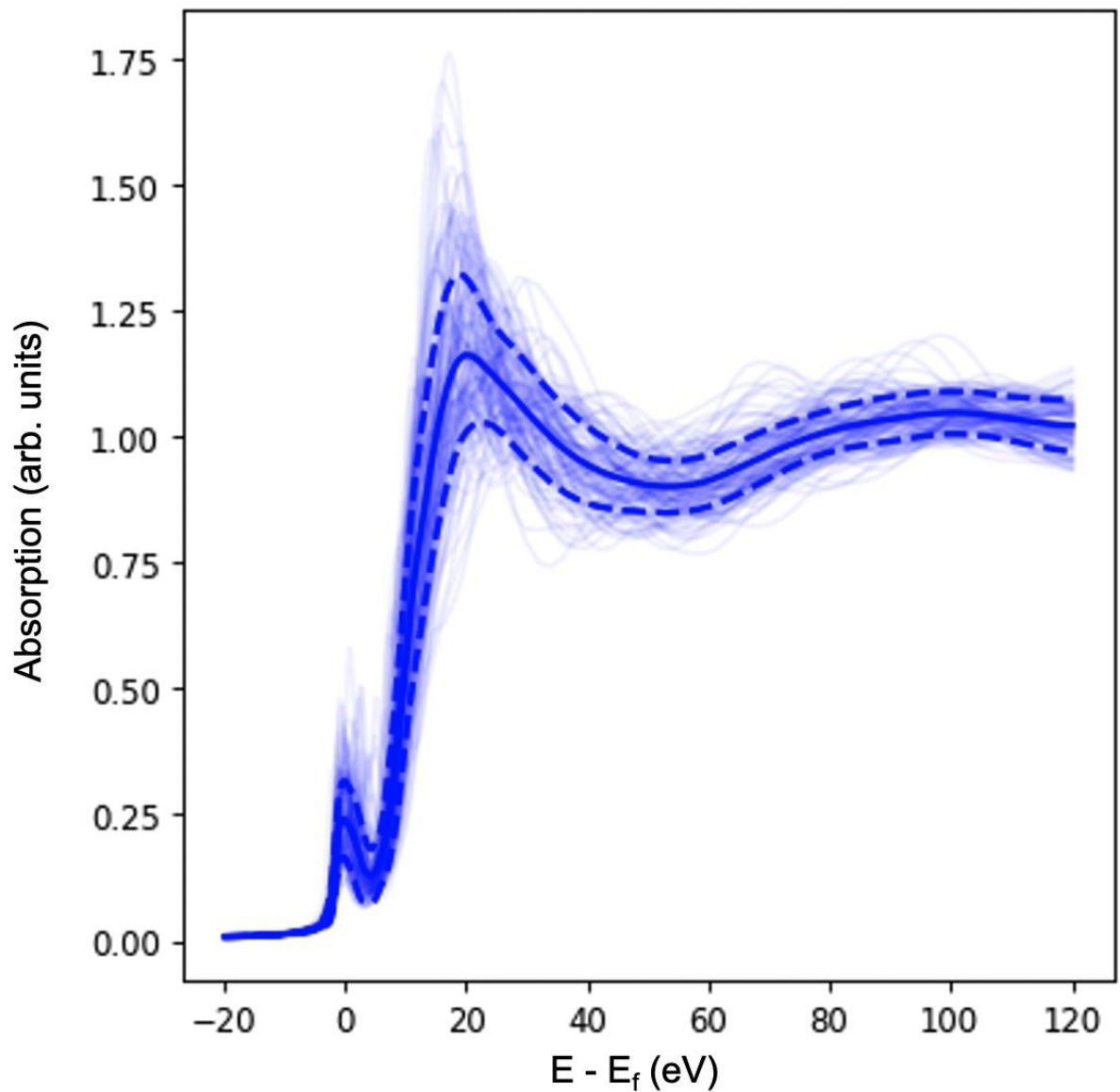

**Supplementary Fig. 10. Computed absorption coefficient.** Computed absorption coefficient of each Fe absorber (thin blue lines) in the cell in dense molten $Fe_{0.6}Mg_{0.4}CO_3$ simulated using DFT-MD simulations, resulting average (blue continuous line), and the standard deviation envelope (blue dashed lines). The energy scale is relative to the Fermi level.

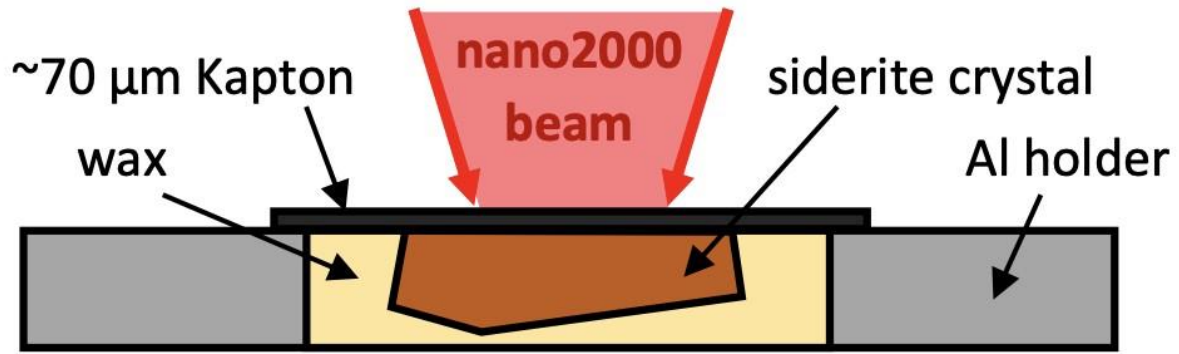

**Supplementary Fig. 11. Sample configuration of the recovery shots performed at the nano2000 facility of the LULI laboratory.** The shock compressed samples were recovered to perform post-shot analysis.